\documentclass[twocolumn]{aastex701}
\usepackage[version=4]{mhchem}

\newcommand{\KBD}{\citetalias{hainline2024a}}

\begin{document}
\defcitealias{hainline2024a}{H24} 
\defcitealias{hainline2026}{H26} 

\title{JWST/NIRSpec Spectra for Three Ultracool Brown Dwarfs Detected in Extragalactic Surveys: Further Evidence for Phosphine Absorption}

\author[orcid=0000-0003-4565-8239,sname='North America']{Kevin N.\ Hainline}
\affiliation{Steward Observatory, University of Arizona, 933 N. Cherry Avenue, Tucson, AZ 85721, USA}
\email[show]{kevinhainline@arizona.edu}  

\author[orcid=0000-0002-6721-1844]{Samuel A. Beiler}
\affiliation{School of Physics, Trinity College Dublin, University of Dublin, College Green, Dublin 2, Ireland}
\email[]{}  

\author[orcid=0000-0002-6523-9536]{Adam J. Burgasser}
\affiliation{Department of Astronomy and Astrophysics, University of California, San Diego, La Jolla, CA, USA.}
\email[]{}

\author[orcid=0009-0003-9152-3252]{Evan S. Chen}
\affiliation{Monta Vista High School, Cupertino, CA, USA}
\email[]{}  

\author[orcid=0000-0003-4337-6211,sname='North America']{Jakob M.\ Helton}
\affiliation{The Department of Astronomy \& Astrophysics, The Pennsylvania State University, 525 Davey Lab, University Park, PA 16802}
\email[]{}  

\author[orcid=0000-0002-0834-6140]{Jarron Leisenring}
\affiliation{Steward Observatory, University of Arizona, 933 N. Cherry Avenue, Tucson, AZ 85721, USA}
\email[]{}  

\author[orcid=0000-0002-5500-4602]{Brittany E.\ Miles}
\affiliation{Steward Observatory, University of Arizona, 933 N. Cherry Avenue, Tucson, AZ 85721, USA}
\email[]{}  

\author[orcid=0000-0002-5251-2943]{Mark S. Marley}
\affiliation{Department of Planetary Sciences, Lunar \& Planetary Laboratory, University of Arizona, Tucson, AZ 85721, USA}
\email[]{} 

\author[orcid=0000-0003-1622-1302]{Sagnick Mukherjee}
\affiliation{School of Earth and Space Exploration, Arizona State University, Tempe, AZ, USA}
\email[]{}  

\author[orcid=0000-0002-4735-8224]{Stefi Baum}
\affiliation{Department of Physics and Astronomy, University of Manitoba, Winnipeg, MB R3T 2N2, Canada}
\email[]{} 

\author[orcid=0000-0002-8651-9879]{Andrew J.\ Bunker}
\affiliation{Department of Physics, University of Oxford, Denys Wilkinson Building, Keble Road, Oxford OX1 3RH, UK}
\email[]{}  

\author[0000-0002-6719-380X]{Stefano Carniani}
\affiliation{Scuola Normale Superiore, Piazza dei Cavalieri 7, I-56126 Pisa, Italy}
\email[]{}  

\author[orcid=0000-0003-2388-8172]{Francesco D'Eugenio}
\affiliation{Kavli Institute for Cosmology, University of Cambridge, Madingley Road, Cambridge, CB3 0HA, UK}
\affiliation{Cavendish Laboratory, University of Cambridge, 19 JJ Thomson Avenue, Cambridge, CB3 0HE, UK}
\email[]{}  

\author[orcid=0000-0003-1344-9475]{Eiichi Egami}
\affiliation{Steward Observatory, University of Arizona, 933 N. Cherry Avenue, Tucson, AZ 85721, USA}
\email[]{}  

\author[orcid=0000-0002-3642-2446]{Tobias J.\ Looser}
\affiliation{Center for Astrophysics $|$ Harvard \& Smithsonian, 60 Garden St., Cambridge MA 02138 USA}
\email[]{}  

\author[orcid=0000-0002-5104-8245]{Pierluigi Rinaldi}
\affiliation{Space Telescope Science Institute, 3700 San Martin Drive, Baltimore, Maryland 21218, USA}
\email[]{}  

\author[orcid=0000-0001-9262-9997]{Christopher N. A. Willmer}
\affiliation{Steward Observatory, University of Arizona, 933 N. Cherry Avenue, Tucson, AZ 85721, USA}
\email[]{}  

\begin{abstract}

We present JWST NIRSpec prism spectra for three ultracool brown dwarfs discovered in extragalactic survey data, two from the JWST Advanced Deep Extragalactic Survey (JADES), and one from Public Release IMaging for Extragalactic Research (PRIMER) survey observed as part of the Mirage or Miracle (MoM) program. The spectra for these sources indicate that one is a T6 dwarf (JADES-GS-BD-11, T$_{\mathrm{eff}} = \sim 700$ K) and two are Y0-Y1 dwarfs (JADES-GS-BD-5,  T$_{\mathrm{eff}} = \sim 400$ K, and MoM-239450, T$_{\mathrm{eff}} = \sim 500$ K). Model atmospheric fits with \texttt{NIFTY} to the spectra are consistent with this classification, and indicate that JADES-GS-BD-5 is only $\sim 150$ pc from the Sun, MoM-239450 is $\sim 700 - 800$ pc from the Sun, and JADES-GS-BD-11 is $\sim 1$ kpc from the Sun, with these latter two more distant sources being best fit at sub-solar metallicities. JADES-GS-BD-5 has an observed spectrum with significantly weaker J and H band emission than Y dwarf atmospheric models, potentially indicating the presence of water ice clouds in the brown dwarf. The spectrum for JADES-GS-BD-11 has a feature at 4.3$\mu$m consistent with absorption from the rarely seen phosphine molecule at $2.6\sigma$ confidence. Given the low metallicity for this source ([M/H] $ = -0.7$), our finding supports the theory that detecting phosphine in brown dwarf atmospheres is tied to atmospheric metallicity. JWST/NIRSpec spectroscopy continues to be a powerful tool for understanding the properties of these distant, and very cold brown dwarfs. 

\end{abstract}


\section{Introduction} \label{sec:introduction}

Brown dwarfs are astronomical sources with masses below the $\sim0.075$ M$_{\odot}$ minimum mass for hydrogen fusion \citep{kumar1962,kumar1963,hayashi1963}, placing them at the intersection of low-mass stars and extrasolar planets. At such low masses brown dwarfs continuously cool after their formation, and they serve as excellent tracers of the history of the region of the Milky Way galaxy where they reside. Their low temperatures (M dwarfs have $2000\ \mathrm{K} < T_{\mathrm{eff}} < 3500\ \mathrm{K}$, L dwarfs have $1200\ \mathrm{K} < T_{\mathrm{eff}} < 2000\ \mathrm{K}$, T dwarfs are found at $500\ \mathrm{K} < T_{\mathrm{eff}} < 1200\ \mathrm{K}$ and Y dwarfs are found at T$_{\mathrm{eff}} < 500$ K) result in the emission of most of their flux at near- to mid-infrared wavelengths. After being discovered in the mid-1990s \citep[see][]{kulkarni1997}, the vast majority of brown dwarfs have been identified through wide-area infrared surveys and lie within 100 pc of the Sun, owing to their faint intrinsic fluxes. The subsequent study of these objects has provided key insights into both star formation and planet atmospheric physics.  

Since the launch of the James Webb Space Telescope \citep[JWST][]{gardner2023}, brown dwarf science has undergone a renaissance, with many sources being identified in deep near-IR extragalactic imaging, observations at flux limits that allow for very cold and, excitingly, very \textit{distant} sources to be identified \citep{nonino2023, burgasser2024, langeroodi2023, hainline2024a, chen2025, hainline2026, li2026}. Molecular absorption in the atmospheres of these sources results in a distinct shape of their spectral energy distribution (SED), with flux peaks at $1-2\mu$m, and a broader, often stronger peak at $3 - 5\mu$m. This emission results in colors that allow brown dwarfs to be differentiated from extragalactic sources, although at low effective temperatures there is still ambiguity between the colors of Y dwarfs and candidate ultra-high-redshift ($z \sim 30$) galaxies \citep[see][]{gandolfi2025b, hainline2026}. Obtaining deep near-IR spectroscopy of these objects is therefore crucial for confirming the nature of these cold sources and understanding their atmospheric chemistry, metal abundances, effective temperatures, specific gravities, and distances. Only a limited number ($\lesssim 40$) of brown dwarfs discovered in extragalactic surveys have been spectroscopically confirmed to date \citep{burgasser2024, hainline2024b, tu2025a, tu2025b, morrissey2026, bradac2026}. The SPHEREx mission promises to increase this number significantly, but primarily for brighter T-dwarfs \citep[see][for early samples]{gagne2026, tu2026} and with lower resolution spectrophotometry.

As brown dwarfs cool, their atmospheres become increasingly dominated by molecules, with vertical mixing transporting species from the deep interior into the observable photosphere, including phosphine (PH$_3$), a tracer of this disequilibrium chemistry. Spectroscopic absorption from phosphine, observed in Jupiter and Saturn \citep{bregman1975, ridgway1976, fletcher2009} and long predicted to be observed in giant planets and brown dwarfs due to vertical mixing pulling phosphine from the interiors of these sources \citep{visscher2006}, has been nonetheless difficult to observe. Brown dwarfs provide a unique opportunity to test the same disequilibrium processes seen in Jupiter and Saturn over a much broader range of temperatures, gravities, and metallicities. However, \citet{miles2020} explored the spectra for 7 local brown dwarfs between $250 - 750$ K, and while the sources had evidence for vertical mixing in their atmospheres, only upper limits on the phosphine abundance were measured. Given that exploring the abundance of phosphine requires observations at 4 - 5$\mu$m, it has historically been difficult to observe these faint features from the ground. 

JWST spectroscopy of local brown dwarfs has been crucial in hunting for phosphine: \citet{burgasser2025} presented exciting evidence of phosphine absorption at $\sim 4.3 \mu$m in the nearby T dwarf Wolf 1130c \citep{mace2013} (T$_{\mathrm{eff}} = 621 \pm 9$K, dist = 19 pc), a brown dwarf with a very low iron abundance [Fe/H] = $-0.70 \pm 0.12$ \citep{mace2018, woolf2006}. This discovery was especially notable given that the presence of the molecule was not observed in other deep brown dwarf spectra obtained with JWST/NIRSpec \citep[][]{beiler2024, faherty2024, kiman2026}, although we should note the one-part-per-billion detection for WISE 0855–0714 presented in \citet{rowland2024}. The low atmospheric metallicity of Wolf 1130c is one of the primary reasons that \citet{burgasser2025} theorize they were able to detect phosphine in the spectrum for this object given the significantly reduced strength of the CO$_2$ absorption. Additionally, a recent theoretical exploration of the molecule presented in \citet{Yin2026} indicated that phosphine may condense to form solid metal phosphides in brown dwarf atmospheres, and at low metallicity the availability of such condensing metals is lowered significantly. 

With so few significant detections, our physical understanding of atmospheric phosphine in brown dwarfs remains limited, providing an impetus to search for the molecule across the population of \textit{distant} brown dwarfs, many of which are at low metallicities similar to Wolf 1130c \citep{burgasser2024, hainline2024b, tu2025b, hainline2026}. Phosphine absorption was potentially observed at very low significance in the T-dwarfs UNCOVER-BD-3 \citep{burgasser2024} and JADES-GS-BD-9 \citep{hainline2024b}. However, caution must be taken given how standard models for substellar atmospheres have difficulty in replicating the $4.1 - 4.4\mu$m wavelength range and predict an overabundance of phosphine, and an underabundance of carbon dioxide \citep{beiler2024a}, which have overlapping opacity. 

The combination of low effective temperatures, low metallicities, and the wavelength coverage of JWST/NIRSpec makes distant brown dwarfs discovered in extragalactic surveys uniquely powerful laboratories for studying disequilibrium atmospheric chemistry. Beyond this, spectroscopy of these distant ultracool dwarfs additionally extends the empirical census of T and Y dwarfs into populations that remain poorly sampled in the Solar neighborhood. Each new spectrum obtained provides an important benchmark for atmospheric models at temperatures and metallicities that are still represented by only a handful of objects. To that end, in this paper we present JWST/NIRSpec prism spectroscopy of three ultracool brown dwarfs identified in deep extragalactic imaging surveys: JADES-GS-BD-5, JADES-GS-BD-11 \citep{hainline2024a} and MoM-239450 \citep[GO-5224, PIs Oesch \& Naidu, see ][]{naidu2026}.

We present the observational properties and photometric and spectroscopic data for these sources in Section \ref{sec:observations}. We fit the NIRSpec spectra in this section using an update to the fitting code \texttt{NIFTY} \citep{hainline2026}, and confirm that all three of these sources are brown dwarfs at T$_{\mathrm{eff}} < 700$K, with distances that range between 100 - 800 pc from the Sun, which we present in Section \ref{sec:results}. The observed spectrum of JADES-GS-BD-11 has tentative evidence for phosphine absorption in a low-metallicity ([M/H] $\sim -0.7$) T dwarf 1000 parsecs away, and we discuss this in Section \ref{sec:discussion}. We finally conclude in Section \ref{sec:conclusions}. 

\section{Observations and Data Reduction} \label{sec:observations}

In this section we describe the three brown dwarfs being explored in this study: JADES-GS-BD-5, JADES-GS-BD-11, and MoM-239450. In Section \ref{subsec:sample} we discuss the sources themselves, their original selection, and their photometric properties. In Section \ref{subsec:reduction} we introduce NIRSpec prism spectra for each source and discuss how these data were reduced and extracted. 

\subsection{Sample Properties} \label{subsec:sample}

The three sources considered in this paper were found within extragalactic deep fields observed with JWST/NIRCam. JADES-GS-BD-5 and JADES-GS-BD-11 were discovered by the JWST Advanced Deep Extragalactic Survey \citep[JADES,][]{eisenstein2023} within the GOODS-S region of the sky, and were first discussed in \citet{hainline2024a}. Updated fits to the JADES Data Release 5 \citep{johnson2026, robertson2026} photometry for these two sources were presented in \citet{hainline2026}. MoM-239450 was targeted in the Public Release IMaging for Extragalactic Research (PRIMER) and COSMOS-Web \citep{casey2023} field as part of the Miracle or Mirage \citep[MoM][]{oesch2024} program, which selected the source for spectroscopic follow-up as part of a search for ``black hole stars'' \citep[see ][]{naidu2025}. Briefly, we summarize the photometric properties and results from fits to these data for these sources below. 

JADES-GS-BD-5 (R.A.: 53.084038, DEC: -27.839346, JADES DR5 ID 190413) has photometry consistent with the source being a Y dwarf: it is seen in multiple NIRCam bands, and while it is very faint between $1 - 2.7\mu$m ($<10$ nJy), it quickly increases in brightness beyond that wavelength and then stays extremely bright ($\sim 10^3$ nJy) out through the JWST/MIRI photometric bands to 12.8$\mu$m. \citet{hainline2026} find that the source has T$_{\mathrm{eff}} = 322^{+5}_{-7}$ K, and is only $\sim 70$ pc from Earth, from a fit with the Sonora Elf Owl models \citep{mukherjee2024, wogan2025} using the code \texttt{NIFTY}\footnote{\url{https://github.com/kevinhainline/NIFTY}}. Both \citet{hainline2024a} and \citet{hainline2026} present a proper motion for this source of $50 \pm 20$ mas yr$^{-1}$, which at the predicted distance, corresponds to a transverse velocity of $17 \pm 6$ km s$^{-1}$.

JADES-GS-BD-11 (R.A.: 53.200314, DEC: -27.764126, JADES DR5 ID 140057), on the other hand, has a bluer $1 - 2\mu$m color in line with the source being a T-dwarf, and \citet{hainline2026} derive T$_{\mathrm{eff}} = 644 \pm 37$ K at $\sim 710$ pc from Earth from their \texttt{NIFTY} fit. While this source falls within the JADES MIRI footprint, because of its distance and spectral shape it is not detected, and the final fit is consistent with the upper limits on the long-wavelength MIRI photometry. This source also has measured proper motions from \citet{hainline2024a} and \citet{hainline2026}: $48 \pm 10$ mas yr$^{-1}$, which at the predicted distance, corresponds to a significantly faster transverse velocity than JADES-GS-BD-5, $162 \pm 40$ km s$^{-1}$, the highest across the whole JADES sample. 

MoM-239450 (RA: 150.098156, DEC: 2.212728) is listed in the COSMOS-Web Data Release 1 \citep{shuntov2025} as ID 693499. In this catalog, when using the non-point-spread-function-homogenized aperture photometry, and adopting 0.5$^{\prime\prime}$ diameter aperture fluxes, the source has an F115W flux of $-20.3 \pm 12.0$ nJy, an F150W flux of $-9.51 \pm 9.4$ nJy, an F277W flux of $15.5 \pm 7.5$ nJy, an F444W flux of $76.4 \pm 7.2$ nJy, and a MIRI F770W flux of $12.7 \pm 1.6$ nJy. The other, non-JWST, photometry presented in the catalog serves as upper limits, and we avoid using these data in the fit because of how the unknown proper motion of the source could potentially bias the aperture photometry. When we fit the five NIRCam + MIRI fluxes with \texttt{NIFTY} using the Sonora Elf Owl models following \citet{hainline2026}, we estimate T$_{\mathrm{eff}} = 418^{+115}_{-95}$ K at $\sim 603$ pc, although we note that this fit underpredicts the F277W flux by an order of magnitude. 

\subsection{NIRSpec Spectroscopy and Reduction} \label{subsec:reduction}

\begin{figure*}[t!]
  \centering
  \includegraphics[width=0.8\linewidth]{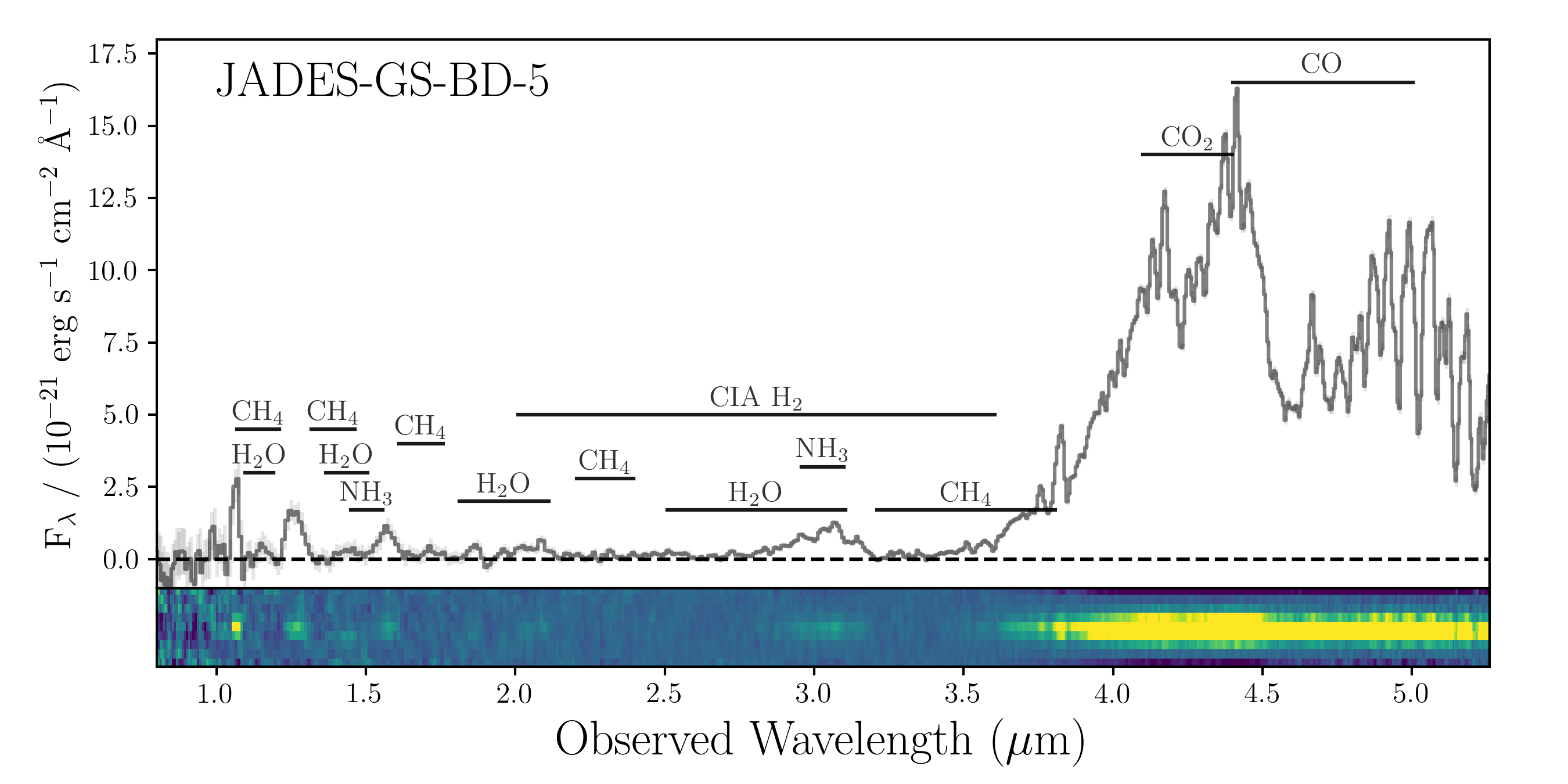}\
  \includegraphics[width=0.8\linewidth]{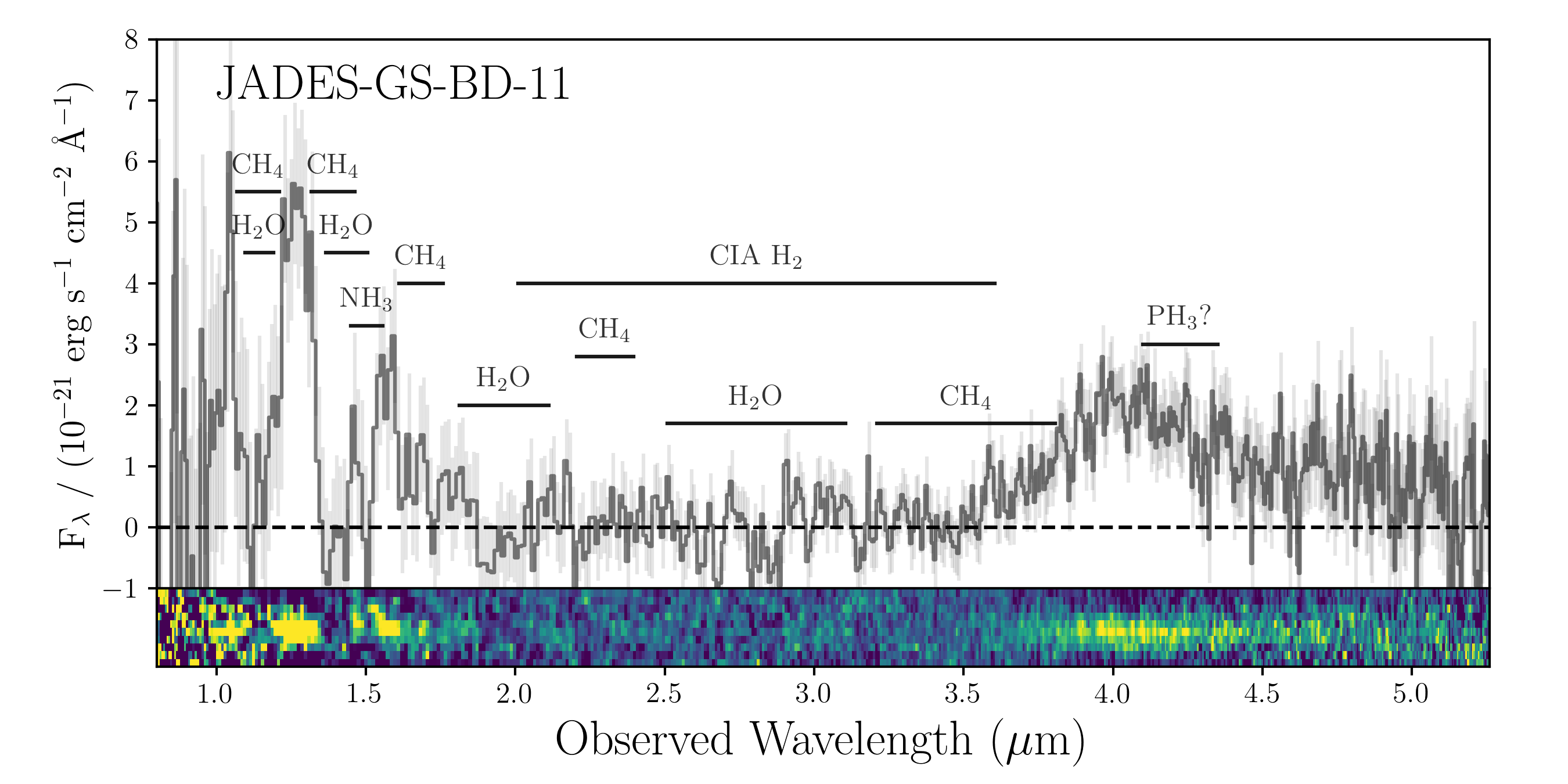}\
  \includegraphics[width=0.8\linewidth]{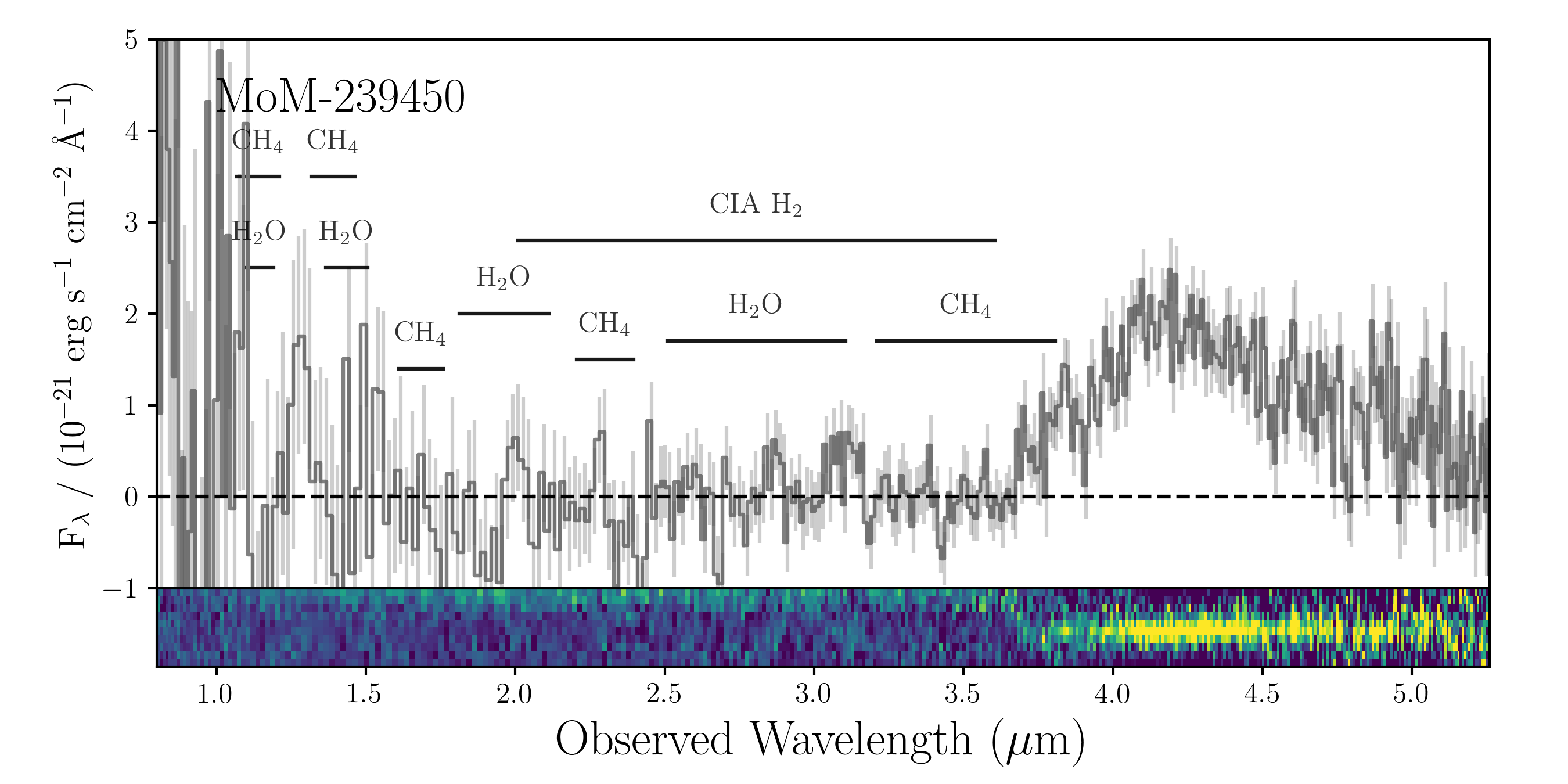}\
  \caption{2D and 1D JWST/NIRSpec PRISM spectra for the three brown dwarfs described in this study. In each 1D spectrum panel, we plot the spectrum in dark grey, and the 1$\sigma$ uncertainty with light grey bars. We also plot common absorption bands from molecules seen in the atmospheres of T and Y dwarfs with black lines. In the 2D spectra, we show the observed fluxes for each source, and vary the normalization to properly demonstrate the significance of the emission and absorption features. (Top) JADES-GS-BD-5, originally selected in \KBD{}. (Middle) JADES-GS-BD-11 originally selected in \KBD{}. (Bottom) MoM-239450.}
  \label{fig:flambda_spectra}
\end{figure*}

The sources described above were observed using the JWST/NIRSpec instrument \citep{jakobsen2022} across three separate observational programs. All three observations used the Micro-Shutter Assembly \citep{ferruit2022} and the prism disperser, covering 0.6--5.5~$\mu$m with resolving power $R\simeq30\text{--}330$ \citetext{\citealt{jakobsen2022}, although the effective resolving power for unresolved sources is higher, e.g., \citealt{degraaff2024,deugenio2025}}.

JADES-GS-BD-5 was observed as part of the JWST Cycle 3 Program Observing All phases of StochastIc Star formation \citep[OASIS, PID 5997, ][]{looser2024}. The source was included on the NIRSpec Multi-Shutter Array (MSA) at the request of the authors, with the position provided after correcting for proper motion. The observations were obtained as part of Observation 3, split between two visits executed between January 5\textsuperscript{th} and 7\textsuperscript{th}, 2025. We used 19 groups per integration, two integrations per exposure, and the NRSIRS2 readout \citep{rauscher2012,rauscher2017}, for a total of 2,800~sec per exposure. We obtained 36 exposures, split in three dithered MSA configurations which were repeated four times each; within each repetition, we used three nodded exposures for accurate background subtraction. The final exposure time was 2,800$\times$36 sec = 100.8~ksec.

JADES-GS-BD-11 was observed as part of the Cycle 4 Program JWST Multi-Cycle Deep Transient Survey in GOODS-S \citep[PID 8060, ][]{egami2025}. Again, the source was included on the MSA at the request of the authors, using coordinates corrected for proper motion to epoch 2026.1. The target was observed as part of Observation 10 on January 13\textsuperscript{th}--14\textsuperscript{th}, 2026. We used the same exposures and readout as was done for the OASIS observations (2,800~sec per exposure) for 9 exposures, split between three dithered MSA configurations; each configuration was observed only once, with three nodded exposures. The final exposure time was 2,800$\times$9=25.2~ksec.

MoM-239450 was included as part of the sources targeted by the Cycle 4 Mirage or Miracle program \citep[MoM, PID 5224, ][]{oesch2024} owing to its extremely red color, and was observed in Observation 3 on April 14\textsuperscript{th}, 2025. The observations used 12 groups per integration, two integrations per exposure, and the NRSIRS2 readout, giving 1,780~sec per exposure. There were 9 exposures, taken with a single MSA configuration that was repeated three times, each time using three nodded exposures. The final exposure time was 16.0~ksec.

Both spectra for JADES-GS-BD-5 and JADES-GS-BD-11 were reduced following the JADES pipeline, as outlined in \citet{bunker2024, deugenio2025, curtislake2025, scholtz2025}. The MoM-239450 spectrum was taken from the DAWN JWST Archive (DJA) v4.4\footnote{\url{https://dawn-cph.github.io/dja/index.html}}, and was reduced using the \texttt{msaexp}\footnote{\url{https://github.com/gbrammer/msaexp}} software, following the procedure outlined in \citet{heintz2024} and \citet{degraaff2025}. For the MoM-239450 spectrum, there is an oversubtraction of the 1D spectrum in the DJA reduction, potentially due to a failed or disobedient shutter in the NIRSpec MSA that is not properly accounted for. To correct this oversubtraction, we follow the procedure discussed for the similar oversubtraction found in the DJA reduction for the source Bullet BD1 in \citet{bradac2026}: we estimated the flux in the DJA 1D spectrum between 1.5 and 3.5$\mu$m and added this flux back as an overall offset ($2.48$ nJy) to the whole spectrum. 

We plot the 2D and 1D NIRSpec spectra for the three sources described in this study in Figure \ref{fig:flambda_spectra}. For JADES-GS-BD-5 and JADES-GS-BD-11, the 1D spectra agree with the JADES DR5 photometry used in \citet{hainline2026} for fitting with \texttt{NIFTY}, and help to confirm that these two sources are indeed brown dwarfs. Both sources show strong molecular absorption bands at $1 - 2\mu$m, and then strong emission at $> 3.5\mu$m. The high signal-to-noise ratio (SNR) for the JADES-GS-BD-5 spectrum at longer near-IR wavelengths includes a variety of absorption bands that are not often observed in Y-dwarfs at this distance. The MoM-239450 spectrum, while also helping to confirm that the source is a brown dwarf, agrees with the NIRCam F444W flux, but is significantly fainter than the F277W catalog aperture photometry, with no significant flux detected at $< 3.5\mu$m. We provide the reduced, pipeline processed 2D and 1D spectra for these three objects, along with fits and ancillary plots, on Zenodo (\url{https://doi.org/10.5281/zenodo.21936947}).

\subsection{Spectral Fits} \label{subsec:fitting}

To explore the properties of these brown dwarfs, we first compared the observed spectra to brown dwarf spectra from the SpeX Prism Library using The SpeX Prism Library Analysis Toolkit \citep[\texttt{SPLAT}, ][]{burgasser2017}. We performed fits to the ``standard'' stars within \texttt{SPLAT} to help confirm the spectral types, although this can only be performed for JADES-GS-BD-5 and JADES-GS-BD-11 as MoM-239450 does not have significant flux at $<2.5\mu$m. Even with this caveat, for BD-5 the temperature of the source and lack of defined Y-dwarf standard spectra within \texttt{SPLAT} limits the use of this method. To overcome this limitation, we also compared our photometry to a sample of Y dwarfs observed using Hubble Space Telescope Wide Field Camera 3 (WFC3) grism spectra and assembled in \citet{schneider2015}, as well as the local ultracool T and Y dwarf JWST/NIRSpec spectra presented in \citet{beiler2024}. We ran simple $\chi^2$ minimization fits to these spectra, with normalization of the model as the only free parameter, and explored the goodness of fit as a function of spectral type. 

For a more detailed understanding of the source properties, we additionally compared the spectroscopy to brown dwarf atmospheric models using multiple fitting codes. First, for the purposes of this study we updated the publicly-available \texttt{NIFTY} software to fit NIRSpec prism spectra in addition to NIRCam and MIRI photometry. \texttt{NIFTY}, introduced in \citet{hainline2026}, is designed to use a grid interpolation of the chosen atmospheric model spectra to perform Markov Chain Monte Carlo (MCMC) fits to the observed photometric points. We simply adapted the code to also allow for a direct fit to a spectrum rather than fitting synthetic model photometry to observed photometry. \texttt{NIFTY} currently works with the Sonora Elf Owl \citep{mukherjee2024, wogan2025}, LOWZ \citep{meisner2021}, and ATMO2020++ \citep{meisner2023}\footnote{\url{https://www.erc-atmo.eu/?page_id=322}} brown dwarf atmospheric models. For JADES-GS-BD-5, model fits using the LOWZ suite of atmospheric models perform poorly as the predicted effective temperature runs up against the low end of the model range (500 K), and so we only fit to the Sonora Elf Owl and ATMO2020++ models for this source. \texttt{NIFTY} performs a Bayesian fit to observed data using the \texttt{emcee} sampler \citep{foremanmackey2013}, with uniform priors on the free parameters for each model. For Sonora Elf Owl and LOWZ, the free parameters are effective temperature, specific gravity, eddy diffusion parameter (log$(K_{zz})$), metallicity ([M/H]), carbon-to-oxygen ratio (C/O), and the overall normalization, which we convert to a distance to the source assuming each is the radius of Jupiter. For ATMO2020++, the free parameters are the same, but without C/O or log$(K_{zz})$. For each model, \texttt{NIFTY} compares the observed fluxes to an interpolated grid created from the original model spectra. For each fit, we restrict the per-pixel flux value to have a SNR $\leq 20$ in order to minimize the effects of imperfect wavelength-dependent flux calibration. 

We also performed fits to the three spectra to a suite of models using \texttt{ucdmcmc}\footnote{\url{https://github.com/aburgasser/ucdmcmc}} \citep{burgasser_2026_ucdmcmc}. \texttt{ucdmcmc} also performs a MCMC fitting, similar to \texttt{NIFTY}, but has a larger selection of brown dwarf models to compare with. For the purposes of this paper, however, we used the same model sets as was done with \texttt{NIFTY} as the majority of the available models used by \texttt{ucdmcmc} do not reach Y-dwarf temperatures. Fits with \texttt{ucdmcmc} allow for the source radius to vary, which differs from \texttt{NIFTY}, such that if the radius predicted by \texttt{ucdmcmc} were smaller than the radius of Jupiter, the source distance would also decrease as compared to what was predicted by \texttt{NIFTY}. We note that \texttt{NIFTY} uses linear interpolation across model grids for comparing to photometry and spectroscopy, and we refer the reader to Appendix A in \citet{tu2024} for a discussion of the uncertainties and use of interpolation in deriving brown dwarf properties.

\section{Results} \label{sec:results}

\subsection{Spectral Types} \label{subsec:spectral_types}

As was discussed in the previous Section, we used fits to the SpeX Prism Library and the Y-dwarfs observed and collected in \citet{schneider2015} and \citet{beiler2024} to estimate the spectral types for the three brown dwarfs in our sample. We will start with JADES-GS-BD-11, where our fit comparing the observed $0.8 - 2.5\mu$m NIRSpec fluxes to the SpeX T dwarf standard sources resulted in a broad minimum in the reduced $\chi^2$ of the fits to the spectrum between T5 and T7 ($\chi^2_{\mathrm{red}} < 0.7$), with a rise in $\chi^2_{\mathrm{red}}$ at earlier and later spectral types. We will consider this source to be a T6 dwarf, based on this fit. 

For JADES-GS-BD-5, the $\chi^2_{\mathrm{red}}$ dropped continually when approaching the latest spectral type available as a standard source in SpeX, but we achieved a slightly better fit when comparing the $0.6 - 1.7\mu$m NIRSpec fluxes to the Y0 dwarfs \citep[the best fit was to the spectrum of WISE 2220-3628, T$_{\mathrm{eff}} = 480 \pm 41$ K, ][]{faherty2024} from \citet{schneider2015}. The best comparison between the JADES-GS-BD-5 spectrum and those presented in \citet{beiler2024} was to that of Y1 dwarf CWISEP J1446-2317 \citep[T$_{\mathrm{eff}} = 366 \pm 14$ K, ][]{beiler2024}, although this spectrum, while matching the $3-5\mu$m flux well, has stronger flux at $1.0 - 1.7\mu$m. Given these comparisons, we consider the source to be a Y0 or Y1 dwarf. 

For MoM-239450, as was discussed earlier, the lack of significant flux at $< 2.7\mu$m made it difficult to compare to standard brown dwarf spectra, but the best-fit to the $0.6 - 1.7\mu$m NIRSpec fluxes was a Y0 brown dwarf \citep[WISE 0825+2805, T$_{\mathrm{eff}} = 359 \pm 3$ K][]{kiman2026} from \citet{schneider2015}. When we compare to the brown dwarfs from \citet{beiler2024}, the best-fit is to the spectrum of the Y0 dwarf WISEJ0734-71 \citep[T$_{\mathrm{eff}} = 493^{+14}_{-24}$ K, ][]{beiler2024}, and we consider MoM-239450 to be a Y0 dwarf as a result.  

\subsection{Spectral Fits} \label{subsec:fits}

\begin{figure*}[t!]
  \centering
  \includegraphics[width=0.82\linewidth]{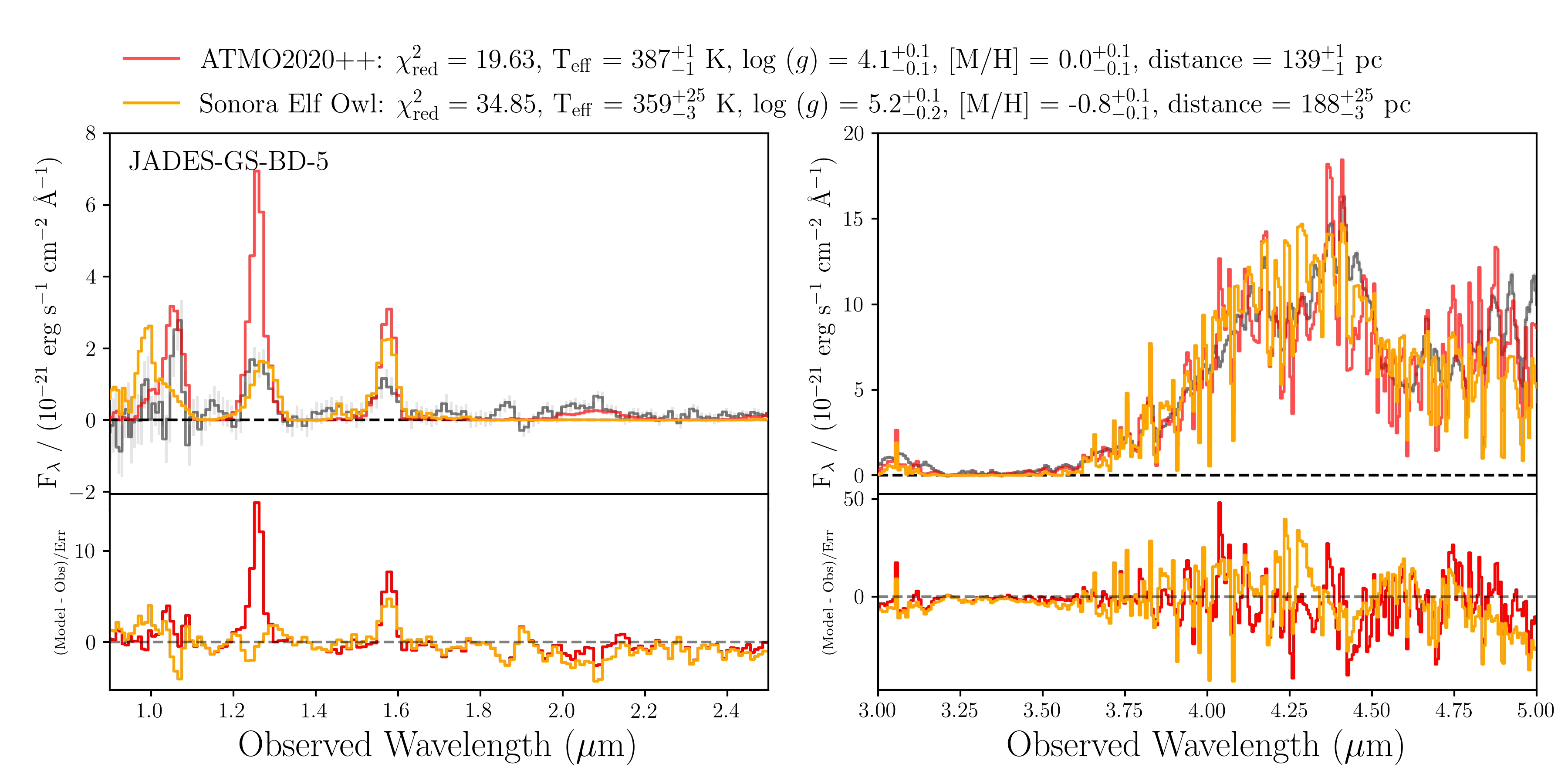}\
  \includegraphics[width=0.82\linewidth]{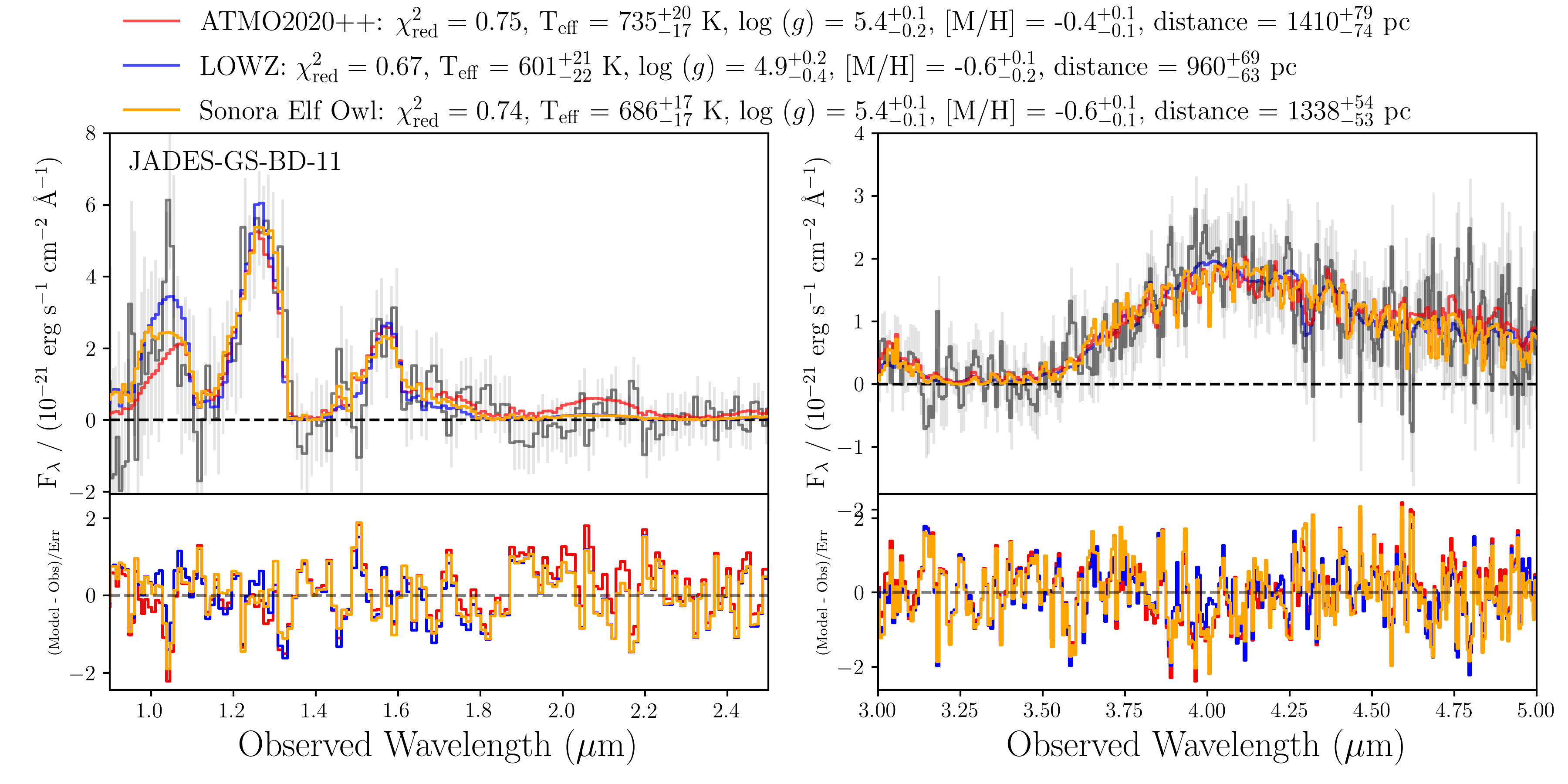}\
  \includegraphics[width=0.82\linewidth]{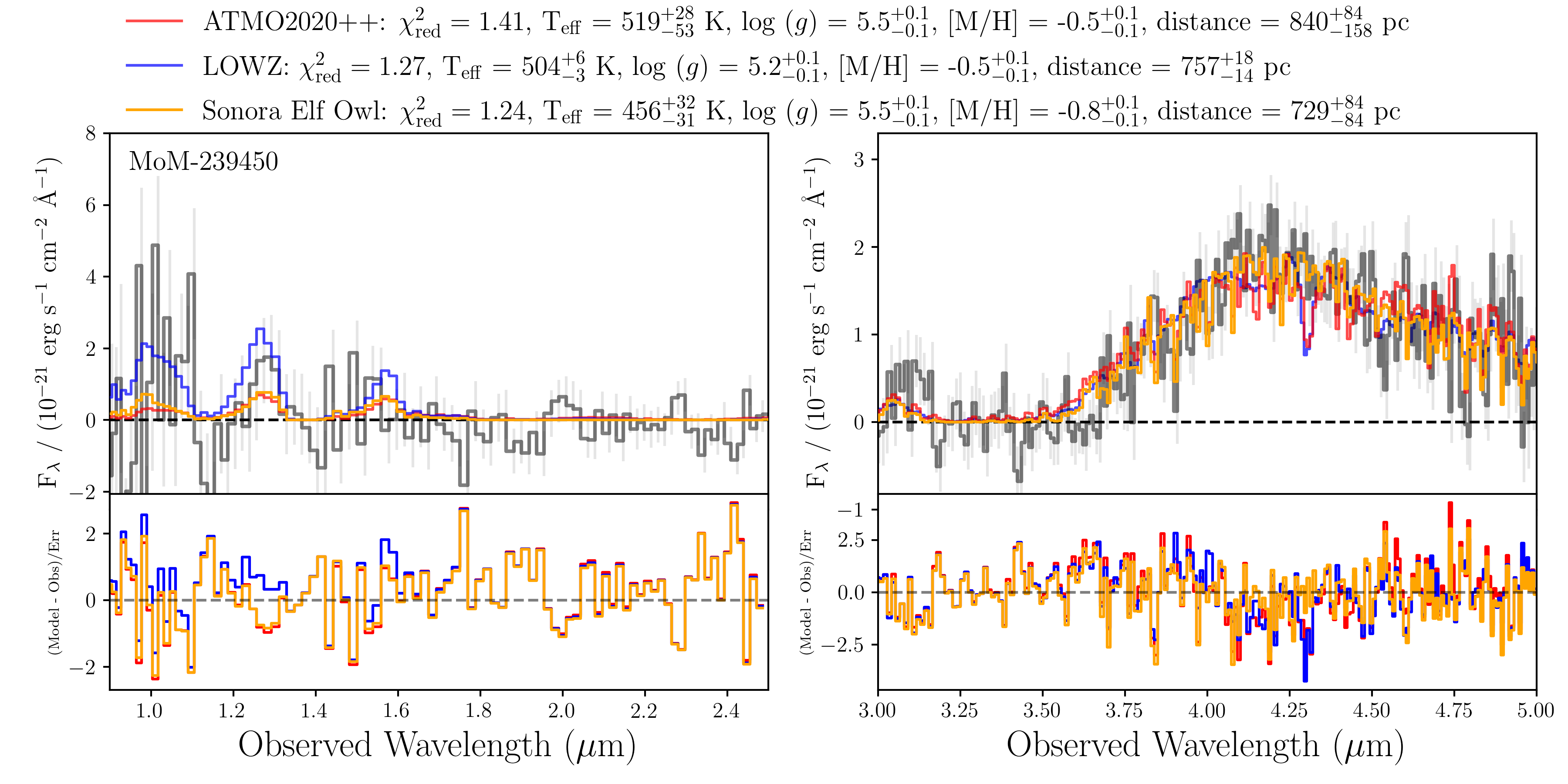}\
  \caption{1D Spectral fits to brown dwarf atmospheric models from \texttt{NIFTY} \citep{hainline2026}. For each source we plot the observed spectrum in grey, and show the fits with colored lines (red: ATMO2020++, \citet{meisner2023}; orange: Sonora Elf Owl, \citet{wogan2025}; blue: LOWZ, \citet{meisner2021}). We plot residuals to each model in the lower panels. We note that the uncertainties presented should be treated as lower limits, as they do not account for underlying systematic errors in the fitting. (Top) JADES-GS-BD-5. (Middle) JADES-GS-BD-11. (Bottom) MoM-239450.}
  \label{fig:sed_fit_fits}
\end{figure*}

In Figure \ref{fig:sed_fit_fits} we plot the best-fit \texttt{NIFTY} SEDs for the models compared to the observed spectra, with fit residuals plotted below, focusing on the $0.9 - 2.5\mu$m wavelength range on the left column, and the $3.0 - 5.0\mu$m emission peak on the right column. Above the plots we show the best-fit parameters for each model, and as a reminder, we do not fit JADES-GS-BD-5 with the LOWZ model atmospheres given the lower limit (500 K) on these model effective temperatures. Overall, the fits for JADES-GS-BD-11 and MoM-239450 are both quite good, while JADES-GS-BD-5 has a significant model vs. data mismatch at the short-wavelength end. For all three, the derived temperatures are in agreement with the spectral types we estimated in the previous subsection, indicating that JADES-GS-BD-5 has an effective temperature of $\sim 380$K and a distance of only $140 - 190$ pc, JADES-GS-BD-11 has an effective temperature of $600 - 700$K and a distance of $1.0 - 1.4$ kpc, and MoM-239450 has an effective temperature of $450 - 500$K with a distance of $700 - 850$ pc. We present the derived parameters for the sources in Table \ref{tab:derived_parameters}.

While the ATMO2020++ and Sonora Elf Owl models largely agree with the flux for JADES-GS-BD-5 at greater than $3.0 \mu$m, both models overpredict the flux at $1.0 - 1.8\mu$m, as seen in the residuals. The diminished short wavelength flux for this source, specifically at $2.0 - 2.2\mu$m, is driving the significantly sub-solar metallicity fit for this object ([M/H] $= -0.8$) using the Sonora Elf Owl Models. When we compare the JADES-GS-BD-5 spectrum to those for more local brown dwarfs from \citet{beiler2024}, we find that the Y0 dwarf WISE J2209+2711 \citep[T$_{\mathrm{eff}} = 379^{+9}_{-13}$ K, ][]{beiler2024}, which is described as ``YJH peculiar'' in \citet{kothari2026}, has similarly weak emission at these wavelengths. We further discuss the origin of these near-IR flux discrepancies in JADES-GS-BD-5 and other such low-temperature sources in Section \ref{sec:discussion}. 

The residuals and the goodness-of-fit values for the fits to the MoM-239450 spectrum additionally demonstrate the difficulties in properly modeling brown dwarfs at such low temperatures, with departures from the models at both $0.9 - 1.8\mu$m (where the fluxes are overpredicted as compared to the LOWZ models, and underpredicted when compared to the Sonora Elf Owl and ATMO2020++ models), and at $3.5 - 3.8\mu$m, resulting in $\chi_{\mathrm{red}}^2 > 1$. Additionally, the ATMO2020++ and LOWZ models include a PH$_3$ feature that is not observed in the spectrum. This source, given its distance, is best fit with [M/H] $< -0.5$, in line with the metallicities observed for other distant, cold brown dwarfs \citep{hainline2026}. 

\begin{figure}[t!]
  \centering
  \includegraphics[width=1.0\linewidth]{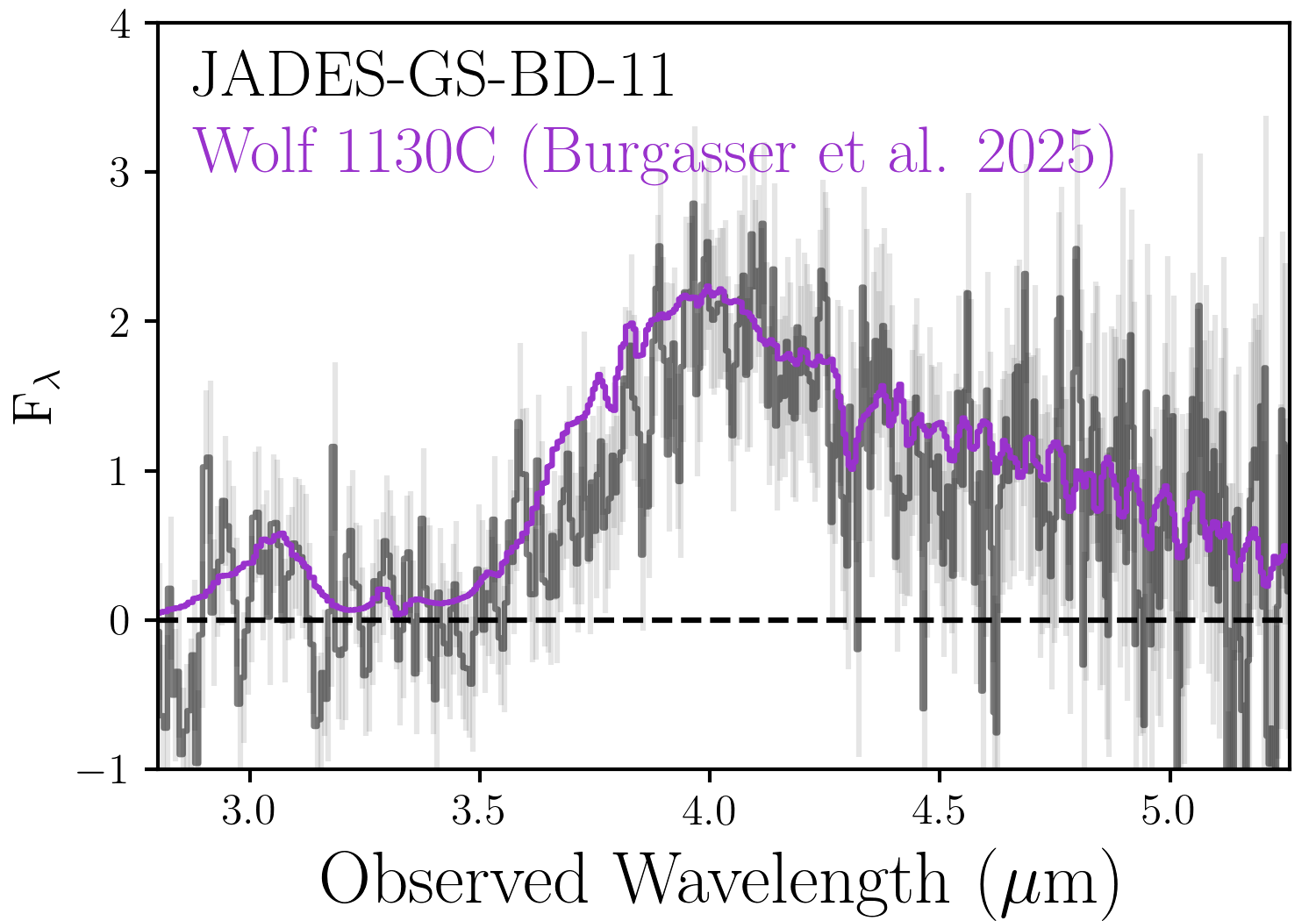}\
  \caption{Comparison of the JADES-GS-BD-11 spectrum at $3.0 - 5.0\mu$m (grey) to the Wolf 1130C spectrum from \citet{burgasser2025} (purple), normalized by matching the median flux at $4.2 - 4.4\mu$m. While at much lower signal-to-noise, the JADES-GS-BD-11 spectrum shows many similarities to the one measured for Wolf 1130C, including potential absorption by H$_2$S at 3.7$\mu$m, NH$_3$ absorption at 3.9$\mu$m, and the phosphine absorption signal at 4.3$\mu$m.}
  \label{fig:comparison_to_wolf1130c}
\end{figure}

JADES-GS-BD-11, however, has fits that are largely consistent with the observed fluxes at all wavelengths, with both the smallest residuals and lowest reduced $\chi^2$ of the three sources we fit in this study. The source is best fit with models at [M/H] $< -0.4$, and we find that JADES-GS-BD-11 is quite similar to Wolf 1130C, which has T$_{\mathrm{eff}} = 621 \pm 9$ K and [M/H] $= -0.5$ \citep{burgasser2025}; we plot both spectra in Figure \ref{fig:comparison_to_wolf1130c}. We make this comparison owing to the observed flux absorption at 4.3$\mu$m in agreement with the model fits, and we surmise this may be phosphine. We will further explore this absorption in Section \ref{sec:phosphine_absorption}.

Our fits to the sources using \texttt{ucdmcmc} largely agree, as shown in Table \ref{tab:derived_parameters}. While the values from each fit are often not within the uncertainties reported by the other fits, this is primarily because uncertainties are underpredicted when using this fitting procedure \citep[see Appendix A in][for a discussion of the uncertainties and use of linear interpolation across model spectra in deriving brown dwarf properties]{tu2024}. Taking into account there large systematic uncertainties, the values are broadly in agreement. Notably, the primary difference is the derived distances, at least for JADES-GS-BD-11 and MoM-239450, although this is likely due to the previously mentioned fact that radius is a free parameter in \texttt{ucdmcmc}, and it is fixed to the radius of Jupiter for \texttt{NIFTY}. The effective temperatures are consistent, although for the LOWZ models, the temperature for JADES-GS-BD-11 and MoM-239450 is much lower when using \texttt{ucdmcmc} than with \texttt{NIFTY}. These results demonstrate how sensitive the final values are on the model set used, and how the uncertainties are largely underpredicted when using this fitting procedure.

\begin{deluxetable}{lc|cccccccr}
\tabletypesize{\footnotesize}
\tablecolumns{8}
\tablewidth{0pt}
\tablecaption{Best-fit atmospheric model parameters derived with \texttt{NIFTY} and \texttt{ucdmcmc} \label{tab:derived_parameters}}
\tablehead{
\colhead{Object} & \colhead{Code} & \colhead{Model}  &   \colhead{T$_{\mathrm{eff}}$ (K)}  &  \colhead{log(g)} & \colhead{log(K$_{\mathrm{zz}}$)} & \colhead{[M/H]} & \colhead{C/O} & \colhead{distance (pc)} & \colhead{$\chi^2_{\mathrm{red}}$}
}
\startdata
JADES-GS-BD-5\tablenotemark{a} & \texttt{NIFTY} & ATMO2020 & $387^{+1}_{-1}$ & $4.1^{+0.1}_{-0.1}$ & - & $0.0^{+0.1}_{-0.1}$ & - & $139^{+1}_{-1}$ & 19.63 \\
 & \texttt{NIFTY} & Sonora Elf Owl & $359^{+25}_{-3}$ & $5.2^{+0.1}_{-0.2}$ & $9.0^{+0.1}_{-1.0}$ & $-0.8^{+0.1}_{-0.1}$ & $2.5^{+0.1}_{-0.1}$ & $188^{+25}_{-3}$ & 34.85 \\
 & \texttt{ucdmcmc} & ATMO2020 & $403^{+11}_{-2}$ & $4.8^{+0.1}_{-0.2}$ & $7.0^{+0.1}_{-0.1}$ & $-0.1^{+0.2}_{-0.1}$ & $0.0^{+0.1}_{-0.1}$ & $120^{+8}_{-5}$ & 7.36 \\
 & \texttt{ucdmcmc} & Sonora Elf Owl & $374^{+29}_{-24}$ & $3.3^{+0.3}_{-0.2}$ & $7.8^{+0.5}_{-0.6}$ & $-0.8^{+0.2}_{-0.2}$ & $1.2^{+0.3}_{-0.2}$ & $132^{+26}_{-23}$ & 14.96 \\
\hline
JADES-GS-BD-11 & \texttt{NIFTY} & ATMO2020 & $735^{+20}_{-17}$ & $5.4^{+0.1}_{-0.1}$ & - & $-0.4^{+0.1}_{-0.1}$ & - & $1410^{+79}_{-74}$ & 0.75 \\
 & \texttt{NIFTY} & LOWZ & $601^{+21}_{-22}$ & $4.9^{+0.2}_{-0.4}$ & $6.6^{+1.0}_{-1.0}$ & $-0.6^{+0.1}_{-0.2}$ & $0.5^{+0.1}_{-0.1}$ & $960^{+69}_{-63}$ & 0.67 \\
 & \texttt{NIFTY} & Sonora Elf Owl & $686^{+17}_{-17}$ & $5.4^{+0.1}_{-0.1}$ & $4.6^{+1.2}_{-1.6}$ & $-0.6^{+0.1}_{-0.1}$ & $0.9^{+0.2}_{-0.2}$ & $1338^{+54}_{-53}$ & 0.74 \\
 & \texttt{ucdmcmc} & ATMO2020 & $658^{+63}_{-59}$ & $5.2^{+0.3}_{-0.5}$ & $5.3^{+1.3}_{-0.9}$ & $-0.3^{+0.3}_{-0.4}$ & $0.0^{+0.1}_{-0.1}$ & $836^{+180}_{-165}$ & 0.87 \\
 & \texttt{ucdmcmc} & LOWZ & $490^{+55}_{-75}$ & $4.6^{+0.6}_{-0.5}$ & $0.0^{+0.1}_{-0.1}$ & $-0.6^{+0.6}_{-0.7}$ & $0.0^{+0.1}_{-0.1}$ & $556^{+146}_{-143}$ & 0.92 \\
 & \texttt{ucdmcmc} & Sonora Elf Owl & $628^{+147}_{-119}$ & $4.7^{+0.4}_{-1.1}$ & $4.1^{+1.6}_{-1.8}$ & $-0.5^{+0.4}_{-0.3}$ & $0.7^{+0.3}_{-0.2}$ & $813^{+411}_{-221}$ & 0.79 \\
\hline
MoM-239450 & \texttt{NIFTY} & ATMO2020 & $519^{+28}_{-53}$ & $5.5^{+0.1}_{-0.1}$ & - & $-0.4^{+0.1}_{-0.1}$ & - & $840^{+84}_{-158}$ & 1.41 \\
 & \texttt{NIFTY} & LOWZ & $504^{+6}_{-3}$ & $5.2^{+0.1}_{-0.1}$ & $6.6^{+0.7}_{-0.7}$ & $-0.5^{+0.1}_{-0.1}$ & $0.8^{+0.1}_{-0.1}$ & $757^{+18}_{-14}$ & 1.27 \\
 & \texttt{NIFTY} & Sonora Elf Owl & $456^{+32}_{-31}$ & $5.4^{+0.1}_{-0.1}$ & $6.5^{+1.2}_{-1.3}$ & $-0.8^{+0.1}_{-0.1}$ & $2.1^{+0.2}_{-0.3}$ & $729^{+84}_{-84}$ & 1.24 \\
 & \texttt{ucdmcmc} & ATMO2020 & $501^{+141}_{-70}$ & $4.7^{+0.2}_{-0.2}$ & $4.6^{+1.5}_{-0.6}$ & $-0.6^{+0.4}_{-0.3}$ & $0.0^{+0.1}_{-0.1}$ & $553^{+378}_{-138}$ & 1.47 \\
 & \texttt{ucdmcmc} & LOWZ & $436^{+80}_{-64}$ & $4.6^{+0.4}_{-0.4}$ & $0.0^{+0.1}_{-0.1}$ & $-1.1^{+0.9}_{-0.4}$ & $0.0^{+0.1}_{-0.1}$ & $535^{+200}_{-153}$ & 1.59 \\
 & \texttt{ucdmcmc} & Sonora Elf Owl & $428^{+87}_{-96}$ & $4.8^{+0.5}_{-0.7}$ & $8.3^{+0.6}_{-0.7}$ & $-0.8^{+0.3}_{-0.2}$ & $1.8^{+0.4}_{-0.7}$ & $481^{+192}_{-172}$ & 1.30 \\
\enddata
\tablenotetext{a}{This source was not fit with the LOWZ models as they do not extend to T$_{\mathrm{eff}} < 500$K.}
\end{deluxetable}

\subsection{Tentative Phosphine Absorption in JADES-GS-BD-11}\label{sec:phosphine_absorption}

The most intriguing feature in the spectrum for JADES-GS-BD-11 is the absorption in the continuum observed at 4.3 microns. This feature can be seen in the 1D spectrum, but also it is visible in the 2D spectrum in Figure \ref{fig:flambda_spectra}, supporting our conclusion that it is not a noise effect. The absorption profile matches quite well with the absorption in the models in Figure \ref{fig:sed_fit_fits} which are driven by phosphine. All three atmospheric models we compare to with \texttt{NIFTY} have phosphine absorption included; the Sonora Elf Owl models we include are those provided by \citet{beiler2024a}\footnote{Zenodo:doi:10.5281/zenodo.11370830} with phosphine abundance consistent with general disequilibrium chemistry. In all three cases, the absorption profile from $4.1 - 4.4\mu$m is remarkably similar to the theoretical absorption spectrum for phosphine \citep[see Figure 1B of][]{burgasser2025}. 

\begin{figure}[t!]
  \centering
  \includegraphics[width=1.0\linewidth]{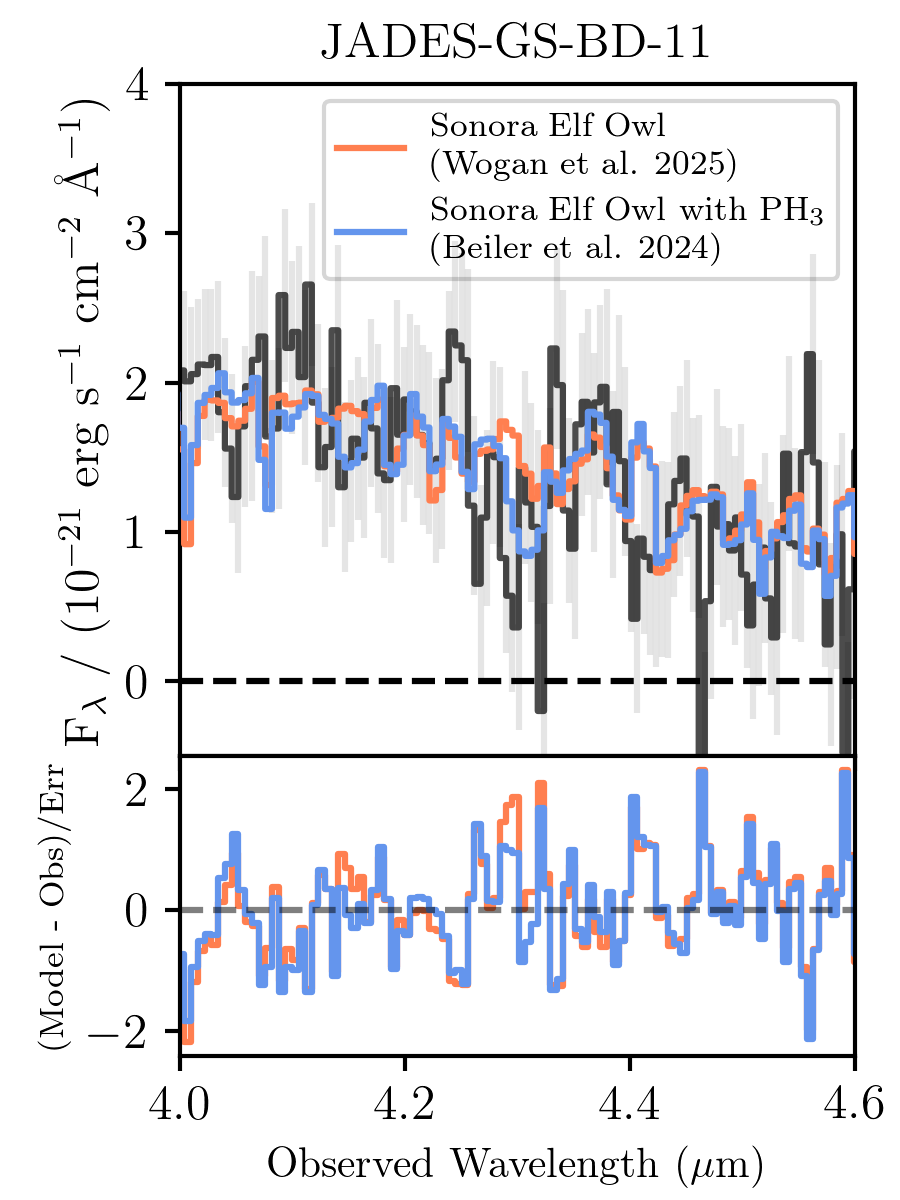}\
  \caption{1D Spectral fits to JADES-GS-BD-11 using brown dwarf atmospheric models from \texttt{NIFTY} with the Sonora Elf Owl 2025 \citep[red, ][]{wogan2025} and 2024 models \citep[blue, ][]{beiler2024a}. The \citet{wogan2025} models do not include the effects of phosphine absorption in the atmosphere, allowing an exploration of the significance of the absorption seen at $\sim4.3\mu$m in JADES-GS-BD-11. The difference in $\chi^2$ for the two fits at 4.2 - 4.4$\mu$m supports the presence of PH$_3$ at 2.6$\sigma$ significance.}
  \label{fig:phosphine_zoom_in}
\end{figure}

To explore this in more detail, we additionally fit the spectrum of JADES-GS-BD-11 with the Sonora Elf Owl models from \citet{beiler2024a}, where PH$_3$ has been included, and we zoom in on $4.0 - 4.6\mu$m and show both fits and the residuals in Figure \ref{fig:phosphine_zoom_in}. While there is significant noise in this wavelength range, the residuals are substantially improved for the fit including phosphine. We quantify this by considering only the $4.2-4.4 \mu$m region: for the fit including phosphine, $\chi_{\mathrm{PH_3}}^2 = 20.0$, whereas the phosphine-free fit yields $\chi_{\mathrm{no\,PH_3}}^2 = 26.8$, indicating a 2.6$\sigma$ preference for the model including phosphine. Importantly, both the phosphine-inclusive and phosphine-free models provide similarly good fits to the 4.7 $\mu$m carbon monoxide (CO) absorption feature, indicating that the improved fit at 4.3 $\mu$m is not achieved at the expense of a poorer fit elsewhere in the spectrum. Moreover, the updated Sonora Elf Owl models already include an enhanced CO$_2$ abundance relative to previous releases, yet the phosphine-free model still fails to reproduce the 4.3 $\mu$m feature. This suggests that the discrepancy cannot be explained solely by an underestimate of the CO$_2$ abundance, strengthening the interpretation that PH$_3$ contributes to the observed absorption.

We should note that this analysis does not include the effects of correlated noise, which would reduce the significance. The most common alternate origin of this feature at this wavelength, CO$_2$, has a highly discrepant absorption profile with a broad minimum at 4.23 $\mu$m that is ruled out given the observed spectrum \citep[see Figure 1C of][]{burgasser2025}. JADES-GS-BD-11 joins UNCOVER-BD-3 \citep{burgasser2024} and JADES-GS-BD-9 \citep{hainline2024b} as distant brown dwarfs with tentative evidence for phosphine absorption, further emphasizing the need for targeted observations of these sources with the SNR necessary to detect the elusive molecule. 

As a further test of this potential PH$_3$, we completed atmospheric retrievals on the spectrum of JADES-GS-BD-11 using \texttt{petitRADTRANS} version 4.0.0a31 (pRT4). pRT4 is latest version of the open-source pRT atmospheric retrieval code. The major reworks that are important to this work are the replacement of the Fortran radiative transfer with a JAX implementation, as well as the inclusion of new retrieval samplers, which provide a significant improvement in the necessary computation time. While this version is currently in beta at the time of writing, it has been comprehensively tested to ensure that results are consistent with the pRT version 3 (Nasedkin et al. 2026, in prep). To test whether the inclusion of PH$_3$ is preferred to model the spectrum of JADES-GS-BD-11, we run two retrievals: one with PH$_3$ included as an opacity source and one without.

For the retrievals run here, we use the correlated-k method for radiative transfer as the NIRSpec spectrum maximum resolving power ($\lambda/\Delta\lambda$=620) is under the limit suggested by pRT's documentation ($\lambda/\Delta\lambda<$1000). For the underlying model, we use a 1D atmospheric model divided in 100 layers ranging from 10$^{-3}$ to 10$^{2.5}$ bar with steps of $0.0\overline{5}$ dex (end point inclusive), which covers the pressure ranges probed by low-resolution NIRSpec data for these cold objects \citep{kothari2026}. The temperature-pressure (T/P) profile is modeled as 7 knot cubic spline evenly spaced in log space between 10$^{-3}$ to 10$^{2.5}$ bar, with each knot retrieved using a uniform prior from 1 to 2000 K. To limit oscillatory wiggles in the T/P profile, a hyperparameter $\gamma$ is used as defined in \citet{Line2015}, with a uniform prior from 0.001 to 1.0. 

Along with the inclusion of PH$_3$ (or lack thereof), we include as opacity sources all the gaseous species that have previous been shown to have clear continuum and absorption features in low-resolution NIRSpec data: H$_2$, He, H$_2$O, CH$_4$, CO, CO$_2$, NH$_3$, Na, and K. H$_2$ and He are not free parameters in the retrieval, as while they make up the bulk composition of brown dwarf atmospheres (74\% and 24\% by mass, respectively) their effects on the spectrum are only seen via Rayleigh scattering and collisionally-induced absorption. The rest of the gas abundances are free parameters assumed to be constant as a function of pressure. All gases have a mass fraction uniform prior from 10$^{-12}$ to 10$^{-0.5}$. The reference for opacities that we use for each molecule are listed in Table \ref{tab:retrievedparameters}, along with a full list of the retrieved free parameters and their priors. We chose to not include clouds in our atmospheric model to lower the number of free parameters and speed up computation time. Clouds have not been shown to be necessary for retrievals to fit the spectra of late T and Y dwarfs (although see the discussion of the JADES-GS-BD-5 spectrum in the next section) and their inclusion do not result in systematic changes to the abundances of most species when included \citep{kothari2026}.

\begin{deluxetable}{llccl}
\tabletypesize{\footnotesize}
\tablecolumns{4}
\tablewidth{0pt}
\tablecaption{Priors and parameters from retrievals of JADES-GS-BD-11 using \texttt{petitRADTRANS} \label{tab:retrievedparameters}}
\tablehead{
\colhead{Parameter}& \colhead{Prior} & \colhead{Median (PH$_3$)} & \colhead{Median (No PH$_3$)}& \colhead{Opacity Reference}}
\startdata
Distance, $d$ (pc)            &$\mathcal{U}(600,1400)$                    & $   1289^{+  54}_{- 160 }$& $   1183^{+  97}_{- 139 }$&$\cdot\cdot\cdot$         \\
Gravity, log (g) [cm s$^{-2}$]&$\mathcal{U}(2.0,5.5)$                     & $   5.12^{+0.11}_{-0.11 }$& $   5.26^{+0.08}_{-0.22 }$&$\cdot\cdot\cdot$         \\
Mass, $M$ ($M_\mathrm{Jup}$)  &$\mathcal{U}(5,75)$                        & $   71.4^{+ 2.5}_{-11.0 }$& $   54.2^{+10.8}_{- 9.6 }$&$\cdot\cdot\cdot$         \\
$T_\mathrm{knot 0}$ (K)       &$\mathcal{U}(1,3000)$                      & $     36^{+  51}_{-  26 }$& $     38^{+  62}_{-  27 }$&$\cdot\cdot\cdot$         \\
$T_\mathrm{knot 1}$ (K)       &$\mathcal{U}(1,3000)$                      & $    103^{+  47}_{-  39 }$& $    110^{+  54}_{-  48 }$&$\cdot\cdot\cdot$         \\
$T_\mathrm{knot 2}$ (K)       &$\mathcal{U}(1,3000)$                      & $    204^{+  48}_{-  44 }$& $    213^{+  55}_{-  53 }$&$\cdot\cdot\cdot$         \\
$T_\mathrm{knot 3}$ (K)       &$\mathcal{U}(1,3000)$                      & $    402^{+  42}_{-  40 }$& $    407^{+  48}_{-  44 }$&$\cdot\cdot\cdot$         \\
$T_\mathrm{knot 4}$ (K)       &$\mathcal{U}(1,3000)$                      & $    741^{+  50}_{-  38 }$& $    743^{+  52}_{-  41 }$&$\cdot\cdot\cdot$         \\
$T_\mathrm{knot 5}$ (K)       &$\mathcal{U}(1,3000)$                      & $   1245^{+  49}_{-  33 }$& $   1240^{+  65}_{-  35 }$&$\cdot\cdot\cdot$         \\
$T_\mathrm{knot 6}$ (K)       &$\mathcal{U}(1,3000)$                      & $   1813^{+  78}_{-  63 }$& $   1816^{+  81}_{-  58 }$&$\cdot\cdot\cdot$         \\
Roughness Penalty, $\gamma$   &$\mathcal{U}(0.001,1)$                     & $   0.93^{+0.05}_{-0.10 }$& $   0.95^{+0.04}_{-0.08 }$&$\cdot\cdot\cdot$         \\
log($f_\mathrm{H_2O}$)        &$\mathcal{U}(-12.0,-0.5)$\tablenotemark{a} & $  -3.11^{+0.15}_{-0.14 }$& $  -3.13^{+0.14}_{-0.16 }$& \citet{Polyansky2018H2O} \\
log($f_\mathrm{CH_4}$)        &$\mathcal{U}(-12.0,-0.5)$\tablenotemark{a} & $  -3.72^{+0.10}_{-0.13 }$& $  -3.78^{+0.10}_{-0.13 }$& \citet{Yurchenko2024CH4} \\
log($f_\mathrm{CO_2}$)        &$\mathcal{U}(-12.0,-0.5)$\tablenotemark{a} &            $<-8.09       $&            $<-7.87$       &\citet{Yurchenko2020CO2} \\ 
log($f_\mathrm{CO}$)          &$\mathcal{U}(-12.0,-0.5)$\tablenotemark{a} & $  -6.11^{+0.39}_{-0.71 }$& $  -6.90^{+0.93}_{-4.37 }$& \citet{Rothman2010CO}    \\
log($f_\mathrm{PH_3}$)        &$\mathcal{U}(-12.0,-0.5)$\tablenotemark{a} & $  -6.82^{+0.23}_{-1.29 }$& $\cdot \cdot \cdot$       & \citet{sousasilva2015PH3}\\
log($f_\mathrm{NH_3}$)        &$\mathcal{U}(-12.0,-0.5)$\tablenotemark{a} &            $<-4.82       $&            $<-4.65$       & \citet{Gordon2022NH3}    \\
log($f_\mathrm{Na}$)          &$\mathcal{U}(-12.0,-0.5)$\tablenotemark{a} &            $<-7.95       $& $  -5.83^{+0.54}_{-6.44 }$& \citet{Allard2019Na}     \\ 
log($f_\mathrm{K}$)           &$\mathcal{U}(-12.0,-0.5)$\tablenotemark{a} & $  -7.46^{+0.22}_{-0.27 }$&            $<-6.98$       & \citet{Molliere2019K}    \\
\enddata
\tablenotetext{a}{The uniform prior for gas abundance is in units of the log of the mass fraction as is the default in pRT, but for ease of comparison we report abundances as the log of the volume mixing rations.}
\end{deluxetable}

The other free parameters we include are log(g), with a uniform prior from 2.0 to 5.5 [cm s$^{-2}$], mass, with a uniform prior from 5 to 75 $M_\mathrm{Jup}$, and distance, with a uniform prior from 600 to 1400 pc. We explore this parameter space using the JAX nested sampler with 400 live points, terminating when the change in the natural log of the evidence falls below 0.01. The low signal-to-noise and resolution of this spectrum results in a quick convergence, occurring at $\sim20,000$ total samples and $\sim4,000$ accepted samples \citep[see][for further explanation details on JAX nested sampling]{albert2023} for both runs. Corner plots showing the 1D posterior probability distributions for each parameter and the covariances shown by the 2D distributions of each pair of parameters can be found at the Zenodo site (\url{https://doi.org/10.5281/zenodo.21936947}), as can the TP profiles with contribution functions. These are summarized in Table \ref{tab:retrievedparameters} as the median and 1$\sigma$ confidence interval that includes the 15.865 and 84.135 percentiles or 3$\sigma$ upper bounds for limited or non-detections. We note that while the abundances were retrieved in units of mass fraction, we report and discuss abundances in terms of volume-mixing ratios for ease of comparison to previous work. 

We find the physical parameters (distance, gravity, and mass) differ by $\sim1\sigma$ between the PH$_3$ and non-PH$_3$ retrievals, with mass showing the largest discrepancy. When compared to the forward models, the retrieved log(g) values are within 1$\sigma$, but are consistently higher. Only the \texttt{NIFTY} ATMO2020++ and Elf Owl model fit distances are consistent with the retrieved distances, while once again the retrieved distances are greater than all those from the forward models.  

Comparing the results from the two retrievals, the temperature profile parameters are nearly identical. As for the chemical abundances, the major species of H$_2$O and CH$_4$ are consistent across runs, and CO is within $\sim$1$\sigma$. While we report the retrieved median abundance for NH$_3$ in Table \ref{tab:retrievedparameters}, both retrievals result in a non-detection due lack of strong features in the near-infrared and the low signal-to-noise of the data. There are discrepancies between the retrieved Na and K abundances, with a strong K detection in the PH$_3$ retrieval and non-detection of Na, and the inverse in the non-PH$_3$ retrieval. While the retrieved abundances are quite discrepant, the model spectrum is not strongly impacted by the difference, as the optical Na and K spectral features are somewhat interchangeable in the low-resolution NIRSpec spectrum of late brown dwarfs, and are often included as a singular parameter in other retrieval frameworks \citep[e.g.][]{Line2015,burningham2021Cloud}. They only impact the spectrum shortward of 1 $\mu$m, a portion of the spectrum where the signal-to-noise is especially limited, via the wings of their respective optical doublets. Considering all of this, we do not find significance in the extreme differences in the retrieved abundances of Na and K.

It is not unexpected that the majority of chemical abundances do not change significantly due to the removal of PH$_3$, as the impact of PH$_3$ on the spectrum is limited to the 3.9$-$4.3 $\mu$m region. The other species with relevant absorption in this region are H$_2$O and CH$_4$, which are strongly constrained elsewhere, and CO$_2$, which in both retrievals only results in a 3$\sigma$ upper bound of $\sim-8$. Such an upper bound is consistent with the abundance that would be expected given the retrieved CO abundance, as there has been shown to be a consistent $\sim2.5$ dex offset between CO and CO$_2$ \citep{kothari2026}.

\begin{figure*}[t!]
  \centering
  \includegraphics[width=\linewidth]{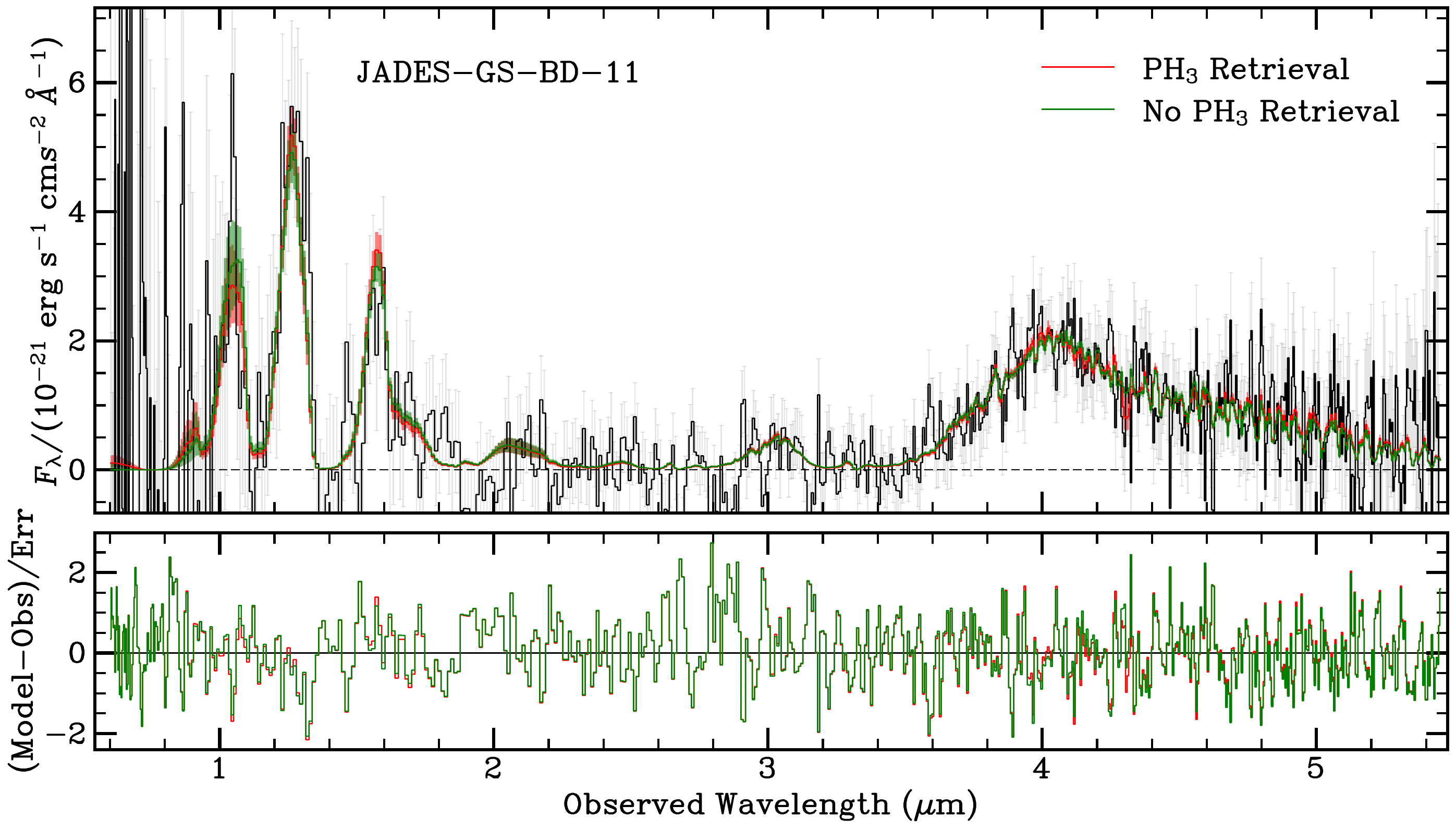}
  \caption{The top panel shows comparisons of the observed spectrum of JADES-GS-BD-11 (black with grey 1$\sigma$ error bars) and the median retrieved spectrum for both the PH$_3$ and non-PH$_3$ \texttt{petitRADTRANS} (red and green, respectively, with 1$\sigma$ central credible intervals shaded around the median). The bottom panel shows the difference between the retrieved and observed spectra, divided by the observed uncertainty. The two retrieved median spectra are generally within 1$\sigma$ of each other everywhere except at 4.3 $\mu$m (the $Q$-branch of the PH$_3$ feature).}
  \label{fig:retrieval_fits}
\end{figure*}

Both retrieval runs result in excellent fits to the data, as seen in Figure \ref{fig:retrieval_fits}. The $\chi^2/\nu$ for the median spectrum of the PH$_3$ model is 0.71, and 0.72 for the median spectrum of the non-PH$_3$ inclusive model. For the PH$_3$ model, the PH$_3$ abundance is sufficient to show a strong feature in the median model spectrum. The retrieved abundance is actually greater than what was found in Wolf 1130c \citep{burgasser2025}, though the uncertainties for JADES-GS-BD-11 are significantly larger. The 1$\sigma$ uncertainties extend down to $-$8.11, which would fall at the edge of what is detectable with even high signal-to-noise NIRSpec PRISM data, as seen in \citet{kothari2026}.

To determine which (or if either) model is preferred, we use two comparison metrics recommended by \citet{thorngren2026}: AIC and BPIC. In both cases the PH$_3$ model is favored ($\Delta \mathrm{AIC}$=$-$5.8 and $\Delta \mathrm{BPIC}=-4.4$). This is enough to point towards an indication of PH$_3$, but not enough to claim a detection based \textit{solely} on these retrieval results.

\section{Discussion} \label{sec:discussion}

The three sources discussed in this survey join the growing number of T and Y dwarfs that have been confirmed within the JWST/NIRCam extragalactic survey data. The spectroscopic confirmation of the Y dwarfs JADES-GS-BD-5 and MoM-239450 are important given the search for ultra-high-redshift ($z > 15$) candidate galaxies. Recently, \citet{bradac2026} presented spectroscopic confirmations of two Y dwarfs found in the field of view of the Bullet Cluster, Bullet BD-1 (T$_{\mathrm{eff}} = 350^{+110}_{-80}$K) and Bullet BD-2 (T$_{\mathrm{eff}} = 410^{+110}_{-50}$K), which were initially chosen as F356W and F277W Lyman-$\alpha$ dropout galaxies at a photometric redshift $z_{\mathrm{phot}} > 30$. The authors measure proper motions for the pair of brown dwarfs: $49 \pm 8$ mas yr$^{-1}$ for Bullet-BD-1 and $24 \pm 3$ mas yr$^{-1}$ for Bullet-BD-2; both values are similar to what is measured for the sources from JADES and presented in \citet{hainline2026}. By comparison, Bullet BD-1 has a spectrum that is similar to JADES-GS-BD-5, but as the latter source is found three to four times farther away, and the exposure time for the JADES-GS-BD-5 spectrum is almost 30 times longer, the spectrum we present here has a higher signal-to-noise. Similarly, Bullet-BD-2 has a spectrum that is comparable to what we present here for MoM-239450, where flux is primarily observed at $> 3.8\mu$m.

In \citet{gandolfi2025b}, the authors present the source Capotauro observed in the CEERS field, which they claim could be a $z \simeq 32$ galaxy, or an ultracool brown dwarf. As discussed in the Section 6.2 of \citet{hainline2026}, the latter scenario is more likely, given that the photometric properties of Capotauro are very similar to what would be expected for JADES-GS-BD-5 if it were observed at a farther distance. The recent analysis in \citet{liu2026} of the source using NIRCam imaging taken 3.5 years following the initial discovery shows that the source has an observed proper motion ($37.6^{+5.5}_{-5.6}$ mas yr$^{-1}$), confirming that it is indeed a Y-dwarf (T$_{\mathrm{eff}} \sim 350$K, at $730 \pm 110$ pc from the Sun). The spectrum we observe for JADES-GS-BD-5 is in agreement with this assessment, and comparable to what is shown in both \citet{gandolfi2025b} and in the updated SEDs presented in \citet{liu2026}. The spectroscopic confirmation of JADES-GS-BD-5, MoM-239450, and now Bullet BD-1 and Bullet BD-2 continue to suggest that care must be taken in searching for ultra high-redshift galaxies given the ubiquitous presence of Y-dwarfs in extragalactic observations, and the large range in space densities for these sources observed in the literature \citep{bradac2026}. 

We have found that the model fits to the JADES-GS-BD-5 spectrum have difficulty replicating the $1 - 1.6\mu$m flux, as shown in the top left panel of Figure \ref{fig:sed_fit_fits}. This wavelength regime probes the deeper, warmer layers of the brown dwarf atmosphere, in contrast to the higher, cooler altitudes probed at $4 - 5\mu$m. The near-infrared flux discrepancy observed in JADES-GS-BD-5 is a known systemic challenge in ultracool substellar modeling, frequently noted in population studies of cold Y dwarfs \citep{leggett2021, fontanive2026}. Characterizing this mismatch has historically been hindered by the lack of observations at wavelengths beyond 2.5$\mu$m for these sources. 

This discrepancy is explicitly highlighted by our MCMC results, where the ATMO2020++ model grid yields a superior overall fit ($\chi^2_{\text{red}} = 19.6$). Both the Sonora Elf Owl and ATMO2020++ models successfully match the complex $\text{CH}_4$ and $\text{CO}_2$ opacities in the 3--5~$\mu$m thermal window while failing to match in slightly different ways shortward of 1.6$\mu$m. The standard cloud-free Sonora Elf Owl model for JADES-GS-BD-5 is forced towards a sub-solar metallicity ($[\mathrm{M/H}] = -0.8$) to statistically reconcile the near-to-mid-infrared color balance. Given that the source's measured proper motion ($50 \pm 20\text{ mas yr}^{-1}$) corresponds to a modest transverse velocity of $17 \pm 6\text{ km s}^{-1}$, which is firmly characteristic of the local (and more likely solar metallicity) thin-disk population, the solar-metallicity fit ($[\text{M/H}] = 0.0$) favored by ATMO2020++ is much more physically plausible. 

While the differences we observe for JADES-GS-BD-5 at short wavelengths could potentially arise from incomplete treatments of alkali opacities, the relatively good fit in the Y-band suggests that potassium profile modeling is not the primary driver of the mismatch. Instead, a more significant effect may be the condensation of water ice clouds in the upper atmosphere. At these low temperatures, such clouds can efficiently scatter and mute the emerging J- and H-band flux peaks while leaving the mid-infrared thermal windows relatively unattenuated \citep{morley2014}. In addition, systemic deficiencies in low-temperature methane and ammonia line lists \citep{zahnle2014, canty2015, tannock2022}, or non-adiabatic alterations to the temperature-pressure profile driven by vertical mixing \citep{tremblin2015,tremblin2019}, remain strong alternative explanations for this structural divergence between the observed data and standard forward-model grids.  This nonphysical set of parameters from the Sonora Elf Owl fits further underscores that the model optimization is attempting to compensate for missing atmospheric physics, such as water-ice cloud opacity or non-adiabatic temperature profiles.

JADES-GS-BD-11, located at a predicted distance of $\sim 1$ kpc from Earth, exhibits a high proper motion (at the \texttt{NIFTY} derived distance using the Sonora Elf Owl models, we find a transverse velocity $v_\mathrm{T} = 234 \pm 51$ km s$^{-1}$) that strongly support an origin within the Milky Way thick disk or inner halo. We can use the proper motion and the derived distance for JADES-GS-BD-11 to explore the orbit of this object around the Milky Way given a range of possible radial velocity values for this brown dwarf. We used the galactic dynamic package \texttt{gala} \citep{adrian_price_whelan_2025_16923466} to estimate a set of orbital parameters for the source across a range of radial velocities between $-450$ to $450$ km s$^{-1}$, and find that the orbits are largely prograde (in the same rotation direction as the Sun), with a higher total orbital energy than the typical cold Milky Way thin disk population \citep{naidu2020}. Our results are more comparable with the kinematically hot tail of the Galactic thick disk, potentially representing an in-situ source dynamically ``splashed'' into the halo \citep[e.g.][]{belokurov2020}. This supports the idea that JADES-GS-BD-11 is an ancient substellar object that likely witnessed the early, chaotic assembly history of the Galactic disk. The sub-solar metallicity we derive from fits to the spectrum ([M/H] $ \lesssim -0.4$) is also in agreement with this hypothesis, given the age of the outer Milky Way. Excitingly, we observe tentative evidence for phosphine absorption in the observed spectrum. While multiple molecular species (including H$_2$O, CH$_4$, CO$_2$, and CO in addition to phosphine) contribute to the complex layout of the $3-5\mu$m spectral region, the dominant contributor at $4.2\mu$m is CO$_2$, which has a very different absorption profile from what we observe in JADES-GS-BD-11 (and seen in Figure \ref{fig:phosphine_zoom_in}). 

Crucially, standard solar metallicity T- and Y-dwarf models are known to underpredict CO$_2$ while overpredicting PH$_3$, creating a notorious pseudo-degeneracy that can mimic or obscure true phosphine features \citep{beiler2024a}. However, because CO$_2$  abundance scales quadratically with atmospheric metallicity, its absorption profile is dramatically suppressed in metal-poor environments. In contrast, the abundance of vertically dredged PH$_3$ drops only linearly with metallicity. It is plausible that the sub-solar metallicity of JADES-GS-BD-11 may have reached a physical threshold where the dominant $\text{CO}_2$ absorption is sufficiently thinned to unmask the underlying phosphine. This matches the exact atmospheric mechanism recently proposed for the metal-poor benchmark brown dwarf Wolf 1130C \citep{burgasser2025}, as well as deep-field targets like UNCOVER-BD-3. While full confirmation of this feature requires higher-resolution spectroscopy to resolve individual vibrational modes or deeper prism integrations to bypass correlated noise problems within the prism spectrum, JADES-GS-BD-11 stands out as a vital, low-metallicity laboratory for testing non-solar disequilibrium chemistry pathways with JWST.

\section{Conclusions} \label{sec:conclusions}

We present JWST/NIRSpec spectroscopy of two Y dwarfs and one T dwarf first identified in extragalactic observations. JADES-GS-BD-5 and JADES-GS-BD-11 were identified in \citet{hainline2024a} and then further discussed in \citet{hainline2026}, and MoM-239450 was found in observations of the COSMOS-PRIMER field and targeted by the Miracle or Mirage program. The spectra for these sources are diverse, with absorption bands from multiple atmospheric molecules and a broad $3 - 5\mu$m peak. We fit these sources both with observed spectra, and with atmospheric models using the updated code \texttt{NIFTY} as well as \texttt{ucdmcmc} and derive effective temperatures, metallicities, specific gravities, and distances for these objects. Both JADES-GS-BD-5 and MoM-239450 have fits indicating they are Y dwarfs, and JADES-GS-BD-11 is best fit at $\sim 600 - 700$K, a late T dwarf. 

At $4.2\mu$m, JADES-GS-BD-11 shows characteristic absorption consistent with phosphine, and we estimate a $2.6\sigma$ confidence level on this detection. This tentative evidence is further supported by atmospheric retrievals run using the \texttt{petitRADTRANS} framework, as the retrieval with PH$_3$ is favored over one without PH$_3$ (though not at a high enough significance to be claimed soley based on the retrieval). If true, this would make it only the second brown dwarf discovered with phosphine in the atmosphere, and would point to a scenario where low metallicity is allowing a detection of atmospheric phosphine in these sources. However, it is currently unknown from such limited samples exactly how common phosphine absorption is in other low-metallicity brown dwarf atmospheres, and whether Wolf 1130C and JADES-GS-BD-11 are anomalies. Given what we can learn about these sources from their near-IR spectra, it is vital for future spectroscopic campaigns that directly target these objects with the exposure times necessary to fully disentangle the effects of molecular absorption from multiple species.  If further phosphine absorption is detected, this will place Wolf 1130C and JADES-GS-BD-11 in context by allowing the study of the temperature range where PH$_3$ is strongest and when the phosphorus abundance becomes trapped in clouds. The empirical relationship for PH$_3$ derived from further observations could potentially be used to benchmark future brown dwarf atmospheric models used on nearby sources. The large population of similar, low metallicity brown dwarfs being found in extragalactic datasets \citep{hainline2024a, hainline2026, chen2025, bradac2026, tu2026} offer a perfect sample for continuing the search for this elusive molecule and updating our understanding of the atmospheres of these cold sources. 

\begin{acknowledgments}
The authors would like to thank Rohan Naidu and Pascal Oesch for the use of the data from the Mirage or Miracle (JWST GO-5224) program, as well as their comments during the writing of this manuscript. This work is based on observations made with the NASA/ESA/CSA James Webb Space Telescope, specifically JADES \citep{rieke2024, eisenstein2023, bunker2024, deugenio2025}. The data were obtained from the Mikulski Archive for Space Telescopes at the Space Telescope Science Institute, which is operated by the Association of Universities for Research in Astronomy, Inc., under NASA contract NAS5-03127 for JWST. KH, JH, JL, and CNAW are supported by the JWST/NIRCam Science Team contract to the University of Arizona, NAS5-02015, and JWST Program 3215. We make use of data from JWST programs 5997, 5224, and 8060. AJB acknowledge funding from the ``FirstGalaxies'' Advanced Grant from the European Research Council (ERC) under the European Union’s Horizon 2020 research and innovation programme (Grant agreement No. 789056). pRT retrievals were performed on the callan system's maintained by the Trinity Centre for High Performance Computing (Research IT). This system is co-funded by the School of Physics and by the Irish Research Council grant award IRCLA/2022/3788.
\end{acknowledgments}

\facilities{HST(ACS), JWST(NIRCam, NIRSpec, and MIRI)}

\software{astropy \citep{astropy2013,astropy2018}, \texttt{NIFTY} \citep{hainline2026}, \texttt{ucdmcmc} \citep{burgasser_2026_ucdmcmc}, \texttt{gala} \citep{gala}, \texttt{petitRADTRANS4} \citep[][ Nasedkin et al. in prep]{Molliere2019K,2024JOSS....9.5875N}, scipy \citep{scipy2020}, numpy \citep{numpy2020}}

\bibliography{ms}{}
\bibliographystyle{aasjournalv7}

\end{document}